\documentclass[10pt,a4paper,twoside]{article}
\usepackage[english]{babel}
\usepackage[latin1]{inputenc}
\usepackage{amssymb}
\usepackage{mathrsfs}
\usepackage{amsmath}
\usepackage[dvips]{graphicx}
\usepackage{epsfig,bm}
\usepackage{graphicx}
\usepackage{rotate}
\usepackage{vmargin}

\usepackage{amsfonts}
\usepackage{graphics}
\usepackage{theorem}

\usepackage{color}
\usepackage{setspace}
\usepackage{amscd}
\usepackage{pstricks}
\usepackage{lscape}
\usepackage{url}
\usepackage[percent]{overpic}
\usepackage{subcaption}

\title{Evaluating performance of neural codes in model neural communication networks}
\author{Chris G. Antonopoulos$^{1}$, Ezequiel Bianco-Martinez$^{2}$ and Murilo S. Baptista$^{3}$}

\date{\today}

\begin{document}

\maketitle

\begin{center}
$^{1}$Department of Mathematical Sciences, University of Essex, Wivenhoe Park, UK\\
$^{2}$Data Science Studio - IBM Netherlands, Amsterdam, The Netherlands\\
$^{3}$Department of Physics (ICSMB), University of Aberdeen, SUPA, Aberdeen, UK

\par\end{center}

\begin{abstract}
Information needs to be appropriately encoded to be reliably transmitted over physical media. Similarly, neurons have their own codes to convey information in the brain. Even though it is well-known that neurons exchange information using a pool of several protocols of spatio-temporal encodings, the suitability of each code and their performance as a function of network parameters and external stimuli is still one of the great mysteries in neuroscience. This paper sheds light on this by modeling small-size networks of chemically and electrically coupled Hindmarsh-Rose spiking neurons. We focus on a class of temporal and firing-rate codes that result from neurons' membrane-potentials and phases, and quantify numerically their performance estimating the Mutual Information Rate, aka the rate of information exchange. Our results suggest that the firing-rate and interspike-intervals codes are more robust to additive Gaussian white noise. In a network of four interconnected neurons and in the absence of such noise, pairs of neurons that have the largest rate of information exchange using the interspike-intervals and firing-rate codes are not adjacent in the network, whereas spike-timings and phase codes (temporal) promote large rate of information exchange for adjacent neurons. If that result would have been possible to extend to larger neural networks, it would suggest that small microcircuits would preferably exchange information using temporal codes (spike-timings and phase codes), whereas on the macroscopic scale, where there would be typically pairs of neurons not directly connected due to the brain's sparsity, firing-rate and interspike-intervals codes would be the most efficient codes.
\end{abstract}


\section{Introduction}\label{sec_intro}

The main function of the brain is to process and represent information, and mediate decisions, behaviors and cognitive functions. The cerebral cortex is responsible for internal representations, maintained and used in decision making, memory, motor control, perception, and subjective experience. Recent studies have shown that the adult human brain has about $86\times 10^9$ neurons \cite{Herculano-Houzel2012}, which are connected to other neurons via as many as $10^{15}$ synaptic connections. Neurophysiology has shown that single neurons make small and understandable contributions to behavior \cite{Perkeletal1969,Schneidmanetal2006,reimann2017cliques}. However, most behaviors involve large numbers of neurons, which are often organized into brain regions, with nearby neurons having similar response properties, and are distributed over a number of anatomically different structures, such as the brain-stem, cerebellum, and cortex. Within each of these regions, there are different types of neurons with different connectivity-patterns and typical responses to inputs.

The coexistence of segregation and integration in the brain is the origin of neural complexity \cite{Sporns2011}. Connectivity is essential for integrating the actions of individual neurons and for enabling cognitive processes, such as memory, attention, and perception. Neurons form a network of connections and communicate with each other mainly by transmitting action potentials, or spikes. To this end, the mechanism of spike-generation is well understood: spikes generate a change in the membrane potential of the target neuron, and when this potential surpasses a threshold, a spike might be generated \cite{Kandeletal1991}. Brain regions show significant specialization with higher functions such as integration, abstract reasoning and consciousness, all emerging from interactions across distributed functional neural networks.

At the local level, the function of individual neurons is relatively well understood. However, the full understanding of the information processing in networks of spiking neurons, the so-called ``neural code'', is still elusive. A neural code is a system of rules and mechanisms by which a signal carries information, with coding involving various brain structures. It is clear that neurons do not communicate only by the frequency of their spikes (i.e. by a rate code) \cite{Gerstneretal1997}, since part of the information can also be transmitted in the precise timing of individual spikes (i.e. temporal code) \cite{Bohte2004}. Also, it is known that some parts of the brain use rate codes (especially motor systems, matching to the slower muscles) and some use timing codes. In some cases, oscillations are very important (e.g. in sniffing), while in others may not be that much \cite{DiLorenzoetal2013}. There is still a debate as to which neural code is used in which brain region, and how much of the potential timing, information is actually used \cite{DiLorenzoetal2013}. Interactions at different timescales might be related to different types of processing, and thus, understanding information processing requires examining the temporal dynamics among neurons and their networks. Precise spike-timing would allow neurons to communicate more information than with random spikes. Different types of neural coding, including temporal and spatial coding, may also coexist on different time scales \cite{Orametal2002}. The scientific evidence so far supports the argument that we are still lacking full understanding on the codes used by neurons to carry and process information, as well as on which neural code is used in which brain region.

What emerges from the scientific evidence so far suggests that fast systems and responses use fast spike-timings coding. For example, the human visual system has been shown to be capable of performing very fast classification \cite{Thorpeetal1996}, where a participating neuron can fire at most one spike. The speed by which auditory information is decoded, and even the generation of speech also suggest that most crucial neural systems of the human brain operate quite fast. For example, human fingertip sensory neurons were found to support this by demonstrating a remarkable precision in the time-to-first spikes from primary sensory neurons \cite{Johanssonetal2004}. Thus, investigating the fundamental properties of neural coding in spiking neurons may allow for the interpretation of population activity and, for understanding better the limitations and abilities of neural computations.

In this paper, we study neural coding and introduce four neural codes. We quantify and compare the rate of information exchange for each code in small-size networks of chemically and electrically coupled Hindmarsh-Rose (HR) spiking neurons \cite{Baptistaetal2010,Antonopoulosetal2015}. We do not deal with spatial codes, but only with temporal and firing-rate codes. For each neuron in the network, we record the temporal courses of its membrane-potential and phase. We construct a suitable map representation of these variables and compute the rate of information exchange for each pair of neurons, aka the Mutual Information Rate (MIR) \cite{Bianco-Martinezetal2016}, as a function of connectivity and synaptic intensities. We consider the precise spike-timings of neural activity (i.e. a temporal code), the maximum points of the phase of neural activities (i.e. neural phase), considering all oscillatory behaviors with arbitrary amplitude, including the high-frequency spiking and low-frequency bursting oscillations, the interspike intervals, and the firing-rate (i.e. ratio of spiking activity over a specific time interval). For the first three codes, we assume that all measurements are performed with respect to the ticks of a local master ``clock'' \cite{Hopfield1995}, meaning relative to the activity produced by one of the participating neurons in the network. This choice is arbitrary in the sense that the activity of any single neuron in the network can be used as the ``clock''. This allows for the estimated mutual information rates to reflect a measure between ``synchronous'' events that occur within a reasonable short-time window. Thus, our estimations provide the strength with which information is exchanged without any significant time-delay, and therefore reflecting a non-directional, non-causal estimation.

In relation to the estimations for the MIR of the neural codes, it would be possible to refine them, considering finer spatial partitions, for example finer than the binary ones considered in this work. These refinements would correspond to the search for a generating, higher-order, Markov partition \cite{Rubidoetal2018}. However, here, we study whether looking at the codes based on the interspike intervals  would provide one with more information than the instantaneous spike-timings code. This is motivated by the question: in a time-series of events, what does carry more information? A code based on the times between events or a code based on the precise times of the occurrence of the events? We control our estimations by comparing them with a theoretical upper bound for MIR to verify the plausibility of the analysis.

Our main findings are summarized as follows: in the simplest case of a single pair of coupled HR spiking neurons, we find that they exchange the largest amount of information per unit of time when the neural code is based on the precise spike-timings. If observable (additive Gaussian white) noise is present, firing-rates are able to exchange larger rates of information than those based on temporal codes and together with the interspike-intervals code are the most robust to noise. In the case of four chemically and electrically coupled HR neurons as in Fig. \ref{fig_results_4neurons}, the largest rate of information exchange can be attributed to the neural codes of the maximum points of the phases (mod 2$\pi$, i.e. to a code dependent on the period of neurons' oscillations) and of the interspike intervals. Surprisingly, pairs of neurons with the largest rate of information exchange using the interspike-intervals and the firing-rate codes are not adjacent in the network, with the spike-timings and phase codes (temporal) promoting large rate of information exchange for adjacent neurons in the network. The latter is also backed by the results in Fig. \ref{fig_results_4neurons_reviewer_question} where connectivity (chemical and electrical connections) is swapped. These results provide evidence for the non-local character of firing-rate codes and local character of temporal codes in models of modular dynamical networks of spiking neurons. When neurons form a multiplex network of 20 HR neurons arranged in two equal-size modules in a bottleneck configuration, communication between pairs of neurons in the two modules is mostly efficient when using either the spike-timings or the maximum points of their phases codes.

\section{Materials and Methods}\label{online_methods}

\subsection{The Hindmarsh-Rose Neural Model}\label{subsection_Hindmarsh-Rose_Model_for_Brain_Dynamics}

We simulate the dynamics of each ``neuron'' by a single Hindmarsh-Rose neuron system. Namely, following \cite{Baptistaetal2010,Antonopoulosetal2015}, we endow each node (i.e. neuron) in the network with the dynamics \cite{Hindmarshetal1984}
\begin{flalign}\label{HR_model_1neuron}
 \dot{p}&=q-ap^3+bp^2-n+I_{ext}\nonumber,\\
 \dot{q}&=c-dp^2-q,\\
 \dot{n}&=r[s(p-p_0)-n]\nonumber,
 \end{flalign}
where $p$ is the membrane potential, $q$ the fast ion current (either Na$^{+}$ or K$^{+}$), and $n$ the slow ion current (for example Ca$^{2+}$). The parameters $a$, $b$, $c$, $d$, which model the function of the fast ion channels, and $s$, $p_0$ are given by $a=1$, $b=3$, $c=1$, $d=5$, $s=4$ and $p_0=-8/5$, respectively. Parameter $r$, which modulates the slow ion channels of the system, is set to $0.005$, and the external current $I_{ext}$ that enters each neuron is fixed to 3.25. For simplicity, all neurons are submitted to the same external current $I_{ext}$. For these values, each neuron can exhibit chaotic behavior and the solution to $p(t)$ exhibits typical multi-scale chaos characterized by spiking and bursting activity, which is consistent with the membrane potential observed in experiments made with single neurons \textit{in vitro} \cite{Hindmarshetal1984}.

We couple the HR system (\ref{HR_model_1neuron}) and create an undirected dynamical network (DN) of $N_n$ neurons connected by electrical (linear diffusive) and chemical (nonlinear) synapses \cite{Baptistaetal2010}
\begin{flalign}\label{HR_model_Nneurons}
\dot{p}_i&=q_i-a p_i^3+bp_i^2-n_i - g_n(p_i-V_{syn})\sum_{j=1}^{N_n}\mathbf{B}_{ij}S(p_j)\nonumber\\&-g_l\sum_{j=1}^{N_n}\mathbf{G}_{ij}H(p_j)+I_{ext}\nonumber,\\
 \dot{q}_i&=c-dp_i^2-q_i,\\
 \dot{n}_i&=r[s(p_i-p_0)-n_i],\nonumber\\
 \dot{\phi}_i&=\frac{\dot{q}_i p_i-\dot{p}_i q_i}{p_i^2+q_i^2},\;i=1,\ldots,N_n\nonumber,
\end{flalign}
where $\dot{\phi}_i$ is the instantaneous angular frequency of the $i$-th neuron \cite{Pereiraetal2007A,Pereiraetal2007B}, $\phi_i$ is the phase defined by the fast variables $(p_i,q_i)$ of the $i$-th neuron, $H(p)=p$ and \cite{Baptistaetal2010}
\begin{equation}\label{sigmoid function}
S(p)=\frac{1}{1+e^{-\lambda(p-\theta_{syn})}}.
\end{equation}
Our work intents to study the transmission of information in models of small-size neural networks by treating them as communication systems, for which we measure and evaluate the rates at which information is exchanged among neurons. We are also not interested in considering realistic biological models for the function $S$. Instead, we consider a biologically inspired function $S$ of a sigmoid type as in Eq. (\ref{sigmoid function}) \cite{Baptistaetal2010}. The remaining parameters $\theta_{syn}=-0.25$, $\lambda=10$, and $V_{syn}=2$ are chosen so as to yield an excitatory DN \cite{Baptistaetal2010}. The synaptic coupling behaves as a short delta-function and carries other features required for synaptic coupling. Particularly, $V_{syn}$ can be tuned to reproduce excitatory or inhibitory behavior, and $S(p)$ has $\theta_{syn}$ to allow for the disconnection of pre-synaptic neurons that have not reached an activation level. For $\lambda=10$, $S(p)$ is a continuous, sigmoid function that behaves similarly to a ``binary process'', either 0 or 1, a fundamental property necessary to use in analytical works when networks with neurons connected simultaneously by electrical and chemical means become synchronous \cite{Baptistaetal2010}. Naturally, this is mimicking the democratic fashion with which chemical synapses behave, where a large community of activated pre-synaptic neurons needs to be activated to induce a relevant response in the post-synaptic neuron(s). This further allows us to study the system knowing the domain of parameters for which we can obtain oscillatory behavior. The choices for the couplings and network topologies in this work are purely abstract, not guided by realistic physiological reasons, and so does time $t$ in Eqs. (\ref{HR_model_1neuron}) and (\ref{HR_model_Nneurons}). The parameters $g_n$ and $g_l$ denote the coupling strength of the chemical and electrical synapses, respectively. The chemical coupling is nonlinear and its functionality is described by the sigmoid function $S(p)$, which acts as a continuous mechanism for the activation and deactivation of the chemical synapses. For the chosen parameters, $|p_i|<2$, with $(p_i-V_{syn})$ being always negative for excitatory networks. If two neurons are connected via an excitatory synapse, then if the pre-synaptic neuron spikes, it might trigger the post-synaptic neuron to spike. We adopt only excitatory chemical synapses here. $\mathbf{G}$ accounts for the way neurons are electrically (diffusively) coupled and is represented by a Laplacian matrix \cite{Baptistaetal2010}
\begin{equation}
\mathbf{G}=\mathbf{K}-\mathbf{A},
\end{equation}
where $\mathbf{A}$ is the binary adjacency matrix of the electrical connections and $\mathbf{K}$ the degree identity matrix of $\mathbf{A}$, leading to $\sum_{j=1}^{N_n}\mathbf{G}_{ij}=0$ as $\mathbf{G}_{ii}=\mathbf{K}_{ii}$ and $\mathbf{G}_{ij}=-\mathbf{A}_{ij}$ for $i\neq j$. By binary we mean that if there is a connection between two neurons, then the entry of the matrix is 1, otherwise it is 0. $\mathbf{B}$ is a binary adjacency matrix and describes how neurons are chemically connected \cite{Baptistaetal2010} and, therefore, its diagonal elements are equal to 0, thus $\sum_{j=1}^{N_n}\mathbf{B}_{ij}=k_i$, where $k_i$ is the degree of the $i$-th neuron. $k_i$ represents the number of chemical links that neuron $i$ receives from all other $j$ neurons in the network. A positive off-diagonal value in both matrices in row $i$ and column $j$ means that neuron $i$ perturbs neuron $j$ with an intensity given by $g_l\mathbf{G}_{ij}$ (electrical diffusive coupling) or by $g_n\mathbf{B}_{ij}$ (chemical excitatory coupling). Therefore, the adjacency matrices $\mathbf{C}$ are given by
\begin{equation}
\mathbf{C}=\mathbf{A}+\mathbf{B}.
\end{equation}
For each neuron $i$, we use the following initial conditions: $p_i=-1.30784489+\eta^r_i$, $q_i=-7.32183132+\eta^r_i$, $n_i=3.35299859+\eta^r_i$ and $\phi_i=0$, where $\eta^r_i$ is a uniformly distributed random number in $[0,0.5]$ for all $i=1,\ldots,N_n$ (see \cite{Antonopoulosetal2015} for details). These initial conditions place the trajectory on the attractor of the dynamics quickly, reducing thus the computational time in the simulations.

\subsection{Numerical Simulations and Upper Bound for MIR} \label{sbnsd}

We have integrated numerically Eqs. (\ref{HR_model_Nneurons}) using Euler's first order method with time-step $\delta t=0.01$ to reduce the numerical complexity and CPU time to feasible levels. A preliminary comparison for trajectories computed for the same parameters (i.e. $\delta t$, initial conditions, etc.) using integration methods of order 2, 3 and 4 (e.g. the Runge-Kutta method) produced similar results. The numerical integration of Eqs. (\ref{HR_model_Nneurons}) was performed for a total integration time of $t_f=10^7$ units and the computation of the various quantities were computed after a transient time $t_t=300$ to make sure that orbits have converged to an attractor of the dynamics. Thus, the sample size used in the estimation of the MIR for the various neural codes is large enough and amounts to 999,970,000 data points (excluding the transient period that corresponds to the first 30000 points).

After Shannon's pioneering work \cite{Shannon1948} on information, it became clear \cite{Borstetal1999,Wibraletal2014} that it is a very useful and important concept as it can measure the amount of uncertainty an observer has about a random event and thus provides a measure of how unpredictable it is. Another concept related to Shannon entropy that can characterize random complex systems is Mutual Information (MI) \cite{Shannon1948}, a measure of how much uncertainty one has about a state variable after observing another state variable in the system. In  \cite{Baptistaetal2012}, the authors have derived an upper bound for the MIR between two nodes or groups of nodes of a complex dynamical network that depends on the two largest Lyapunov exponents $l_1$ and $l_2$ of the subspace formed by the dynamics of the pair of nodes. Particularly, they have shown that
\begin{equation}
\mbox{MIR}\leq I_c=l_1-l_2,\quad l_1\geq l_2,
\end{equation}
where $l_1$, $l_2$ are the two finite-time and -size Lyapunov exponents calculated in the 2-dimensional observation space of the dynamics of the pair of nodes \cite{Baptistaetal2012,Antonopoulosetal2014}. Typically, $l_1$, $l_2$ approach the two largest Lyapunov exponents $\lambda_1$, $\lambda_2$ of the dynamics of the DN (\ref{HR_model_Nneurons}) if the network is connected and the time to calculate $l_1$, $l_2$ is sufficiently small \cite{Baptistaetal2012}. $I_c$ is an upper bound for MIR between any pair of neurons in the network, where MIR is measured in the 2-dimensional observation space. It can be estimated using mainly two approaches: (i) the expansion rates between any pair of neurons, taking the maximal value among all measurements \cite{Baptistaetal2012}, which is tricky and difficult to compute \cite{Baptistaetal2012} and (ii) the two largest positive Lyapunov exponents $\lambda_1$, $\lambda_2$ of the DN. Here, we use the second approach, since the equations of motion of the dynamics are available (Eqs. (\ref{HR_model_Nneurons})) and the full spectrum of Lyapunov exponents can be calculated \cite{Benettin1980}. Particularly, we estimate $I_c$ by $I_c=\lambda_1 - \lambda_2$ (assuming that $l_1 \approx \lambda_1$ and $l_2 \approx \lambda_2$) which will stand for an approximation to the upper bound for the MIR in the network. The phase spaces of the dynamical systems associated to the DNs are multi-dimensional and thus, estimating an upper bound for MIR using $\lambda_1$ and $\lambda_2$, reduces considerably the complexity of the calculations. Besides, parameter changes that cause positive or negative changes in MIR are reflected in the upper bound $I_c$ with the same proportion \cite{Baptistaetal2012}.

\subsection{Estimation of MIR for Maps}\label{subsection_computation_MIR_maps}

There is a huge body of work on the estimation of MI in dynamical systems and in neuroscience, for example \cite{Baptistaetal2012,Kraskovetal2004,Spornsetal2004,Paulusetal2001,Paninski2003,Steueretal2002,Dimitrovetal2001,Wibraletal2014,Rubidoetal2018}.
In this work, we follow the method introduced recently in \cite{CausalityChaos2018}, to estimate MI between pairs of time-series $X(t)$ and $Y(t)$. This methodology is the same one used in \cite{Rubidoetal2018} to estimate MI using refinements or generating Markov partitions. Below, we explain how we estimate MIR by using the estimated values for MI. Pairs $X(t)$ and $Y(t)$ can represent a mapping of any two variables, as in the neural codes introduced in this work. Particularly, we estimate MI by considering binary symbolic dynamics that encode each time-series $X(t)$ and $Y(t)$ into the symbolic trajectory represented by $(\bf{\alpha}, {\beta})$. $N$ sequentially mapped points of $X(t)$ and $Y(t)$ are encoded into  the symbolic sequences $\alpha=\alpha_1,\alpha_2, \alpha_3, \ldots, \alpha_N$ and $\beta=\beta_1,\beta_2, \beta_3, \ldots, \beta_N$, each composed of $N$ elements. The encoding is done by firstly normalizing the time-series $X(t)$ and $Y(t)$ to fit the unit interval. Both $\alpha_i$ and $\beta_i$ can assume only two values, either ``0'', if the value is smaller than 0.5 , or ``1'', otherwise. 

The Mutual Information, MI$(L)$, between $X(t)$ and $Y(t)$ is thus estimated by the MI between the two symbolic sequences $\alpha$ and $\beta$ by
\begin{eqnarray}
\mbox{MI}_{XY}(L) =\nonumber \\\sum_k \sum_l P(X(L)^{\alpha}_k,Y(L)^{\beta}_l) \log{\frac{P(X(L)^{\alpha}_k,Y(L)^{\beta}_l)}{P(X(L)^{\alpha}_k)P(Y(L)^{\beta}_l)}}, 
\label{mutual-information}
\end{eqnarray}
where $P(X(L)^{\alpha}_k,Y(L)^{\beta}_l)$ is the joint probability between symbolic sequences of length $L$ observed simultaneously in $\alpha$ and $\beta$, and $P(X(L)^{\alpha}_k)$ and $P(X(L)^{\beta}_l)$ are the marginal probabilities of symbolic sequences of length $L$ in the sequences $\alpha$ and $\beta$, respectively. The subindices $k$ and $l$ vary from 1 up to the number of symbolic sequences of different lengths $L$ observed in $\alpha$ and $\beta$, respectively. 

MIR is estimated by the slope of the curve of the MI for symbolic sequences of length $L \in [2,5]$ with respect to $L$, which amounts to grid sizes of smaller and smaller cells as $L$ increases. Consequently, this MIR can be considered as an estimation of the increase of MI per time interval. More details can be found in \cite{CausalityChaos2018,Bianco-Martinezetal2016}.

The $L$ interval considers sequences starting from $L=2$ to $L=5$ bits. The reason is that MI behaves linearly with $L$ in this interval, allowing the calculation of MIR. Particularly, the built-in correlations in the time-series and the fact that the chosen partition is likely not the best possible, suggests the exclusion of $L=1$, since correlations would start appearing for larger symbolic sequences. Also, we do not consider $L>5$ as we would run into numerical problems and would introduce under-sampling effects because of the time-series length. For example, if we would assume $L=5$, the analysis for 2 neurons would effectively deal with symbolic sequences of length $2L=10$ and there would be $2^{10}$ different sequences of length $L=10$. A significant trajectory length would then have to be larger than $10\;2^{10}=10240$ trajectory points. Due to the ergodic property of chaotic systems, the probability of observing a given symbolic sequence of length $L$ in one neuron and another of the same length in another neuron is equivalent to the probability of finding trajectory points in a cell of the phase space. The larger $L$ is, the smaller the cell is, and thus, it contains more information about the state of the pair of neurons.

Our MIR$_{ii}$ estimator in Subsec. \ref{subsection_MIR_ii} for the interspike intervals is similar to the work in \cite{Strongetal1998}. In \cite{Strongetal1998}, the authors encode the time-signal by making a time partition, where temporal bins are defined, and a binary encoding is done, by associating 0's to bins without spiking and 1's to bins with spikes.  Our encodings in Sec. \ref{subsection_neural_codes} are based on partitions of the space created by the two time-series $X(t)$ and $Y(t)$. In our approach, we have not sought to maximize MI and search for the generating Markov partition as in \cite{Rubidoetal2018}. However, we have dealt with biasing, when we compare our MIR estimations of the neural codes with $I_c$ estimated by the difference of the two maximal Lyapunov exponents $\lambda_1$ and $\lambda_2$. All our MIR estimations in Sec. \ref{subsection_neural_codes} are bounded by the mathematical upper bound $I_c$ for MIR, except for three cases on which we elaborate later.

It is worth it to note that the parameters and initial conditions in Eqs. (\ref{HR_model_Nneurons}) give rise to chaotic behavior with positive Lyapunov exponents. Thus, chaos is responsible for generating the probabilities necessary for the estimation of MIR in the 2-dimensional spaces of the data from the encoding of the trajectories of pairs of neurons \cite{CausalityChaos2018}. Chaotic behavior in turn gives rise to uncertainty and production of information. Information is then transmitted through the various nodes in the neural network through the electrical and chemical connections (see Eqs. (\ref{HR_model_Nneurons})).

\subsection{Neural Codes}\label{subsection_neural_codes}

Here, we introduce four neural codes and their methodologies to quantify the rate of information exchange between pairs of neurons.

The first uses the spike-timings of neural activity (temporal code), the second the maximum points of the phase of neural activities (neural phase), the third the interspike intervals and the fourth, the firing-rates (ratio of spiking activity over a specific time interval). For the first three, we assume that all recordings are done with respect to the ticks of a local master ``clock'' \cite{Hopfield1995}, relative to the activity produced by a single neuron. This choice can be arbitrary in the sense that the activity of any single neuron can be used. The purpose is to obtain MIR values that can be interpreted as being the current rate of information exchanged between any two neurons, and not any time-delay mutual information. For the estimation of the MIR of the neural codes, we integrate numerically the system of Eqs. (\ref{HR_model_Nneurons}) as discussed in Subsec. \ref{sbnsd} to obtain the numerical solutions to $p_i,q_i, n_i, \phi_i,\; i=1,\ldots,N_n$ as a function of time. We then use these solutions (time-series) to construct the pairs of time-series $X(t)$ and $Y(t)$ to estimate the MIR for each particular neural code as explained below.

Our coupling and topology choices are abstract and inspired by current research in multilayer networks \cite{Baptistaetal2016,Sevilla-Escobozaetal2016,Leyvaetal2017,Antonopoulosetal2017}. We seek to study whether looking at the instantaneous spike-timings provides less information than the codes based on the interspike intervals. Particularly, our study is motivated by the question: having a time-series of events, what does carry more information? A code based on the times between events, or a code based on the exact times of the occurrence of the events?

\subsubsection{Neural Code Based on spike-timings: MIR$_{st}$}\label{subsection_MIR_st}

Here, we explain how we estimate the amount of information exchanged per unit of time between neurons $i,j$ based on the spike-timings of the first neuron, MIR$_{st}$, where $st$ stands for spike-timings. Particularly, we assume that the first neuron plays the role of the ``clock'' and we record $p_i,p_j$ from Eqs. (\ref{HR_model_Nneurons}) at times when $p_1$ of the first neuron attains its local maxima. This allows us to construct a time-series of events $X_i(t),Y_j(t)$ by transforming the continuous dynamics of variables $p_i,p_j$ into a time-series of discrete-time spike events $X_i,Y_j$. We then use $X_i,Y_j$ to compute the rate of information exchanged between neurons $i$ and $j$ as explained in Subsec. \ref{subsection_computation_MIR_maps}. We divide the rate of information exchanged by the mean of the interspike times of the spike activity of the first neuron. We call this quantity, MIR$_{st}$ of the pair of neurons $i,j$.

\subsubsection{Neural Code Based on Phase: MIR$_{m\phi}$}\label{subsection_MIR_mphi}

Next, we explain how we estimate the amount of information exchanged per unit of time between neurons $i,j$ based on the maximum points of the time evolution of the phase variables $\phi_i,\phi_j$, what we denote by MIR$_{m\phi}$,  where $m\phi$ stands for maximum phase $\phi$. We assume that the first neuron plays the role of the ``clock'' and record in time $\Phi_i\equiv\mbox{mod}(\phi_i,2\pi)$ and $\Phi_j\equiv\mbox{mod}(\phi_j,2\pi)$ from Eqs. (\ref{HR_model_Nneurons}) at times when $\Phi_1$ of the first neuron attains local maxima as a function of time $t$. This allows us to construct a time-series of events $X_i(t),Y_j(t)$ by transforming the continuous dynamics of the phase variables of both neurons into a time-series of discrete time events $X_i,Y_j$. We then use $X_i,Y_j$ to estimate the rate of information exchanged between neurons $i,j$. We  divide the rate of information exchanged by the mean of the time intervals for $\Phi_1$ of the first neuron to attain its local maxima. We call this quantity, MIR$_{m\phi}$ of pair $i,j$.

\subsubsection{Neural Code Based on interspike Intervals: MIR$_{ii}$}\label{subsection_MIR_ii}

Here, we show how we estimate the amount of information exchanged per unit of time between neurons $i,j$ based on the interspike intervals of their $p_i,p_j$ variables, denoted by MIR$_{ii}$, where $ii$ stands for interspike intervals. Each neuron can produce a series of different interspike intervals in the course of time. When measuring MI between two interacting systems, we need to specify two correlated relevant events occurring at roughly the same time, if no time-delays are to be considered. These events need to match, i.e., one event happening for one neuron needs to be correlated to one event happening to the other neuron. In order to relate two such time-series with matched pairs of events, we introduce the notion of a relative ``clock''. An interspike interval in neuron $i$ will be matched to the interspike interval of neuron $j$, if neuron $j$ spikes after neuron $i$. Notice that by doing this, we neglect several spikes happening for both neurons, however, we produce a discrete two-dimensional variable that is meaningfully correlated, and therefore, producing a meaningful MI. Another cumbersome approach would be to find an appropriate time-interval within which the two neurons spike, and then correlate their spike-timings intervals or a method as complicated as to calculate the MI considering interspike intervals occurring at different time-delays. This analysis would be more complicated than the one adopted in this work, and would not be necessary. This allows us to construct a time-series of interspike events $X_i(t),Y_j(t)$ from the continuous trajectories of both neurons. We then use $X_i,Y_j$ to compute the rate of information exchanged between neurons $i$ and $j$, dividing this by the mean of the time intervals constructed as the difference between the spike-timings of neuron $j$ and those of neuron $i$, given that the spike of neuron $j$ occurred after that of neuron $i$. We call this quantity, MIR$_{ii}$ of pair $i,j$.

\subsubsection{Neural Code Based on firing-rates: MIR$_{fr}$}\label{subsection_MIR_fr}

Lastly, we show how we estimate the amount of information exchanged per unit of time between neurons $i,j$ for the firing-rates of the $p_i,p_j$ variables, MIR$_{fr}$, where $fr$ stands for firing-rate. Here, we divide the time window between the first and last recorded spike-timing of neuron $i$ into $1.5\times10^6$ equal-size time windows, and compute the firing-rates for both neurons in these time windows. By firing-rate, we mean the ratio between the number of spikes in a given time interval divided by the length of the time interval. This allows us to construct a time-series of firing-rate events $X_i(t),Y_j(t)$. We then use these time-series to compute the rate of information exchanged between neurons $i,j$, dividing it by the length of the equal-size time windows. We call this quantity, MIR$_{fr}$ of pair $i,j$.

\subsection{The link between interspike-intervals and firing-rate codes}

There is a link between interspike-intervals and firing-rate codes that goes back to Kac's lemma \cite{Kacslemma}, which relates return-time intervals, first Poincar\'e returns of the trajectory recurring to a region in phase space, with the probability measure of the trajectory returning to a region in phase space. The firing rate is calculated by $f=N/t$, where $N$ denotes the number of spikes in the time window $t$. Now, $t=\sum_i^N \tau_i$, where $\tau_i$ represents the first Poincar\'e returns, which could also represent the interspike intervals. Defining the average interspike interval by $\langle\tau\rangle=\sum_i^N \tau_i/N$, one can see that $f=1/\langle\ \tau\rangle$. The last equation relates firing rates with the average spike times, though in a statistical sense.

\section{Results}\label{sec_results}

\subsection{Neural Codes for the Communication of Two Neurons}\label{subsec_2neurons}

\begin{figure*}[!ht]
\centering{
\includegraphics[width=\textwidth,height=17.5cm]{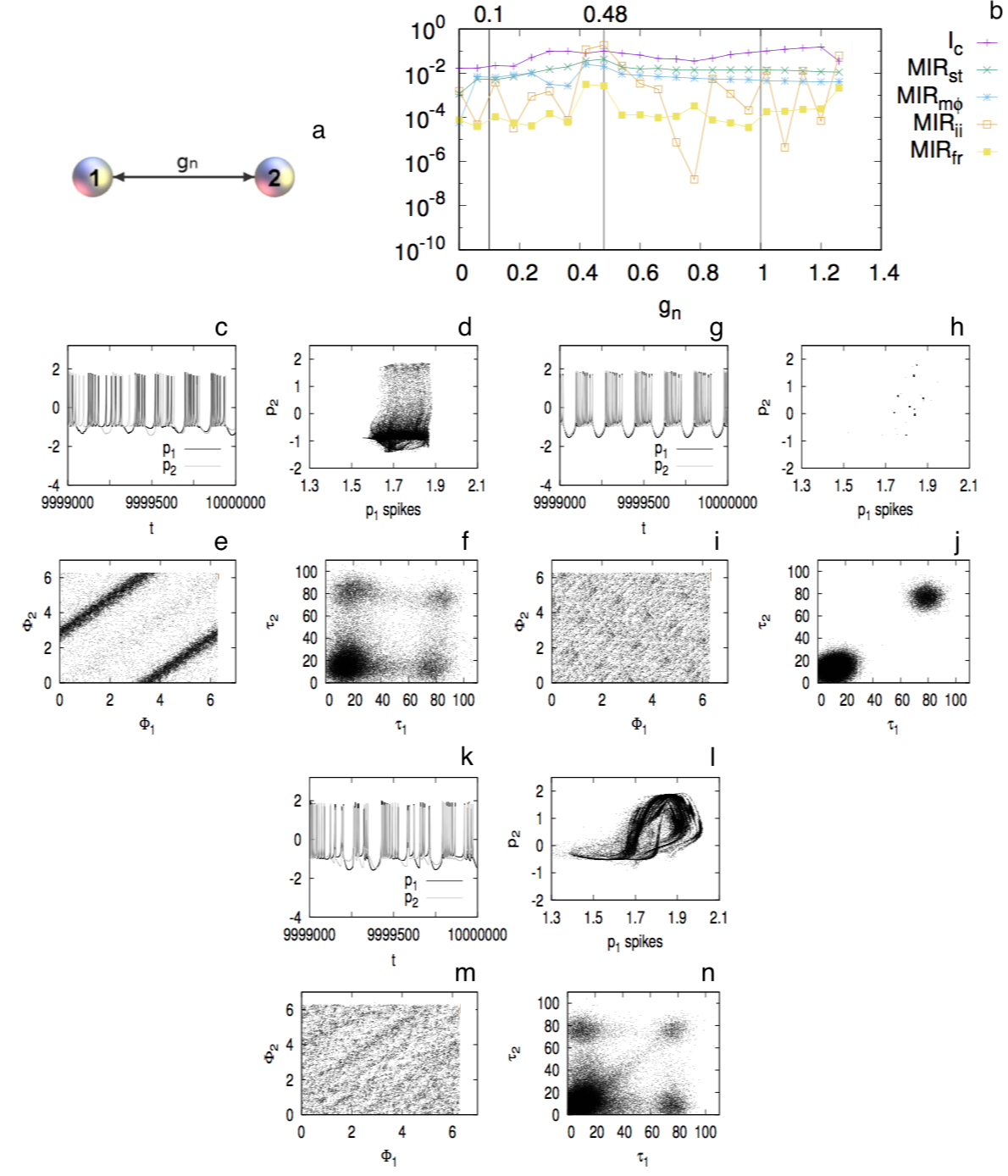}
}
\caption{\textbf{Results for the neural communication channel and the code used between two chemically, bidirectionally connected, non-noisy HR neurons.} Panel a: the pair of chemically connected neurons, where $g_n$ is the strength of the chemical coupling. Panel b: $I_c$, the MIR of spike-timings MIR$_{st}$, MIR of the maxima of the phases MIR$_{m\phi}$, MIR of the interspike intervals MIR$_{ii}$ and MIR of the firing-rates MIR$_{fr}$, respectively. Panels c to f: $p_1,p_2$ as a function of time in panel c, the plane of phase variables $(\Phi_1,\Phi_2)$ (panel d) and, the data used to compute MIR$_{ii}$ (panel f), where $\tau_i,\;i=1,2$ are the interspike intervals of both neurons. Panels g to j: similarly for $g_n=0.48$ and panels k to n for $g_n=1$. In panel b, $g_n=0.1$ that corresponds to a case where MIR$_{m\phi}>$MIR$_{st}$, $g_n=0.48$ to a case where MIR$_{ii}>I_c$, MIR$_{m\phi}$ and MIR$_{st}$, and the case for $g_n=1$ that corresponds to MIR$_{st}>$MIR$_{m\phi}$.}\label{fig_results_2neurons}
\end{figure*}

We study the four neural codes introduced, in the simplest case of a pair of chemically and bidirectionally connected HR neurons (see Fig. \ref{fig_results_2neurons}a), in the absence of noise as we consider its effect in the next section. Our goal is to understand which neural code can maximize the rate of information exchange between the two neurons, considering them as a communication system. We are also interested in finding the chemical coupling strengths $g_n$ this is happening. This is motivated by the question how would two neurons exchange information when disconnected from a network, acting as a single pair. We note that we are not interested in the directionality of the information flow but only in the rate of information exchanged pair-wise. Particularly, in Fig. \ref{fig_results_2neurons}, we calculate the amount of MI per unit of time exchanged between the two neurons, {\em aka} their MIR, for the four neural codes and for different chemical coupling strengths $g_n$.

Before we proceed with a detailed analysis, we summarize the main results of this figure as we increase the synaptic coupling strength from 0.1 (panels c-f), to 0.48 (panels g-j) and to 1 (panels k-n). The time-series of the membrane potential is not synchronous for $g_n=0.1$, it becomes strongly synchronous for $g_n=0.48$, and weakly synchronous for $g_n=1$. The scenario is similar to the amplitudes of the spikes, in panels d, h and l. There is no localization in panel d for $g_n=0.1$, total localization in h for $g_n=0.48$, and partial localization in panel l. In fact, Fig. \ref{fig_results_2neurons}l is the most interesting, as it shows there is a low-dimensional attractor associated to the membrane potential, something usually observed when coupled systems are generalized-synchronous. This scenario changes considerably when observing the behavior of the phases (panels e, i, and m) and the spike-timings (panels f, j, and n). Even though, there is weak phase-synchronization in panel e for $g_n=0.1$, there is no apparent synchronization in panels i and m for $g_n=0.48$ and $g_n=1$, respectively. Figure \ref{fig_results_2neurons}c shows that the membrane potentials ($p_1$ and $p_2$) are mainly asynchronous in time with epochs (time intervals) of synchronicity already visible in this small interval depicted in the panel. This is indeed happening in the full time-series of the numerical simulations (not shown in Fig. \ref{fig_results_2neurons}) and is depicted in Fig. \ref{fig_results_2neurons}e for the phases which are defined in Eqs. (\ref{HR_model_Nneurons}) as a function of $p_1$, $p_2$, $q_1$, $q_2$ and their derivatives. However, the interspike intervals become weakly synchronous in panels f and n, for $g_n=0.1$ and $g_n=1$, respectively, and strongly synchronous in j for $g_n=0.48$. These differences in the intensity of how neurons exchange information will be explored further in the following.

We first see in Fig. \ref{fig_results_2neurons}b that MIR$_{st}$ and MIR$_{m\phi}$ are bigger than MIR$_{ii}$ and MIR$_{fr}$ in certain regions of intermediate and large enough chemical coupling strengths $g_n$. Almost all MIR quantities are smaller than the upper bound for MIR, $I_c$, except for three chemical coupling strengths, ranging from smaller to larger values. This intriguing result is due to the fact that when calculating $I_c$ using the Lyapunov exponents $\lambda_1$ and $\lambda_2$ as discussed previously, $I_c$ is an approximation to the real upper bound for MIR. Consequently, when comparing this upper bound to any lower bound estimations for the MIR, as the neural codes in our work, it might happen these lower bounds be larger than the estimated upper bounds. We note that for $g_n$ values larger than about 1.3, the dynamics becomes quasi-periodic and thus, there is no production of information. The reason is that in such cases, the largest Lyapunov exponent of the dynamics is negative, and consequently there is no chaos, but quasi-periodic behavior which gives rise to predictability, lack of uncertainty and thus no production of information.

In Fig. \ref{fig_results_2neurons}b, we present MIR based on the spike-timings of both neurons (MIR$_{st}$), the MIR for the maximum values of the phase (MIR$_{m\phi}$), the MIR of the interspike intervals (MIR$_{ii}$) and the MIR of the firing-rates (MIR$_{fr}$). We also plot the upper bound for MIR, i.e. $I_c=\lambda_1-\lambda_2$ \cite{Baptistaetal2012}. We focus on three characteristic cases: the first corresponds to the case where MIR$_{m\phi}>$MIR$_{st}$ for chemical coupling strength $g_n=0.1$. The second, to a case where MIR$_{ii}>I_c$, MIR$_{m\phi}$ and MIR$_{st}$ for $g_n=0.48$ (one of the three distinct cases where the computed MIR is bigger than the upper bound $I_c$), and the last one to a case where MIR$_{st}>$MIR$_{m\phi}$ for $g_n=1$. In the first case, the two neurons communicate more efficiently by exchanging larger amounts of information per unit of time using their phases whereas in the third case by exchanging information by the precise spike-timings. In the second case, the two neurons communicate more efficiently by encoding their information in their interspike activity.

To appreciate the performance of the four codes, we first focus on the case of $g_n=0.1$ for which MIR$_{m\phi}>$MIR$_{st}$. We plot in Fig. \ref{fig_results_2neurons}c the time evolution of $p_1,p_2$, the data used to compute MIR$_{st}$ in panel d, the plane of the phase variables of both neurons $(\Phi_1,\Phi_2)$ in panel e, in which the computation of MIR$_{m\phi}$ is based, and in panel f, the data used to compute MIR$_{ii}$, where $\tau_i,\;i=1,2$ are the interspike intervals of both neurons. We observe in panel c that the spike times of both neurons are different. Fig.  \ref{fig_results_2neurons}c shows that the membrane potentials of the two neurons (p$_1$ and p$_2$) are mainly asynchronous in time with epochs (time intervals) of synchronised activity already visible in this small interval depicted in the panel. This is characterized by the two neurons having a phase shift of about $\pi$. The displacement between the phases causes a time shift in the spike trains, thus making time-discrete variables such as the spike-timings asynchronous. The phase reflects the continuous oscillatory behavior of the trajectory, and is connected to the zero Lyapunov exponent. The spike trains reflect its timely character, and is connected to the positive Lyapunov exponent. In fact, both spike  trains and phases are asynchronous. However, when discretization filters out the continuous oscillatory behavior of the trajectory, such as those producing the spike-timings, it is expected that asynchronous behavior is more noticeable from the timing variables due to its connection to the positive Lyapunov exponents.

Particularly, panel d shows that when the first neuron spikes, the second usually remains silent as there is a high density of $p_2$ values around -1, with spikes occurring around $p_2\approx1.9$. This behavior is due to the second neuron which is actually in its quiescent period when the first is spiking. In contrast, when observing the plane of phases in panel e, it becomes apparent that there are two regions of high phase-synchronicity (i.e. stripes of high concentration) and the rest of the region with considerably smaller concentration of phase points. This behavior indicates that the two neurons communicate by chaotically adapting their phases. For the same $g_n$, panel f indicates that the interspike activity of both neurons is well spread in the plane with a high concentration of points occurring close to the origin. Moreover, MIR$_{fr}$ is seen to attain the smaller value with respect to all other quantities.

In panels g to j we study the second case, for $g_n=0.48$, for which MIR$_{ii}>I_c$. We note that this apparent violation comes about because we estimate $I_c$ by the Lyapunov exponents and not by the expansion rates. Since MIR is estimated by a mesh grid of finite resolution, an upper bound for MIR calculated for this grid would require the calculation of expansion rates using the very same grid resolution. $I_c$ estimated by Lyapunov exponents is smaller than the bound estimated by expansion rates (see Supplementary material in \cite{Baptistaetal2012}). Therefore, $I_c$ in this case could not be a true upper bound for MIR. Here, we also observe that MIR$_{st}>$MIR$_{m\phi}$ (see panel b), a result that shows that the two neurons communicate mostly by exchanging information by their precise spike-timings and less by their phases. This can be appreciated in panel g where both $p$ variables attain approximately similar amplitudes during their time evolution. It becomes evident in panel h where the second neuron spikes when the first neuron spikes and that both attain approximately the same amplitudes in their $p$ time-evolution. This behavior is highly localized. In contrast, panel i shows that their phases actually spread all over $[0,2\pi]\times[0,2\pi]$ and that there is no localization of points as it happened for $g_n=0.1$ in which the two neurons communicate by exchanging the largest amount of information per unit of time by their phases. Here, panel j indicates that the interspike activity of both neurons is well localized in two regions with high concentration closer to the origin and on the right upper part of the plot. Moreover, MIR$_{fr}$ is seen to attain the smallest value for this particular chemical coupling strength.

Finally, we focus on the third characteristic case in which MIR$_{st}>$MIR$_{m\phi}$ for $g_n=1$. The situation here is quite different. Indeed, panel k reveals a phenomenon in which the spike times and quiescent periods of both neurons are actually similar. Particularly, panel l reveals that most of the times, either when the first neuron spikes, the second spikes or when the first is in its quiescent period, so is the second, showing a higher density of points in the upper right corner of the plot (spike activity) and a smaller one in its lower left corner (quiescent period). In contrast, the plane of phases in panel m reveals there is no phase synchronization in their activity, as there are no dense regions as in the first case in which MIR$_{m\phi}>$MIR$_{st}$. These results show that the two neurons communicate by their spike-timings, i.e. they use a temporal neural code in which the time of each spike conveys information that is transmitted to other neurons. Lastly, panel n exhibits an interspike activity mostly concentrated in the lower left corner of the plot and less in the other three, a situation completely different to the behavior in panel j of the second case. MIR$_{fr}$ is seen here to attain the smallest value, similarly to the first case.

\subsection{Neural Codes for the Communication Between Two Noisy Neurons}\label{subsec_2noisyneurons}

We now study the same problem in the presence of noise. We consider the effect of additive Gaussian white noise in the performance of the neural codes introduced in Subsec. \ref{subsection_neural_codes}. We want to understand which neural code is more robust to the increase of the noise strength $\sigma$, a case which is more close to realistic neural behavior \cite{Shadlen1994569,RIS_0}. Particularly, in the neural activity of variable $p$ of each neuron, we add white Gaussian noise with standard deviation $\sigma$ to obtain its noisy signal $\bar{p}$:
\begin{equation}\label{Gaussin_noise}
\bar{p}=p+\sigma \mathcal{N}(0,1),
\end{equation}
where $\mathcal{N}(0,1)$ is the Gaussian distribution of zero mean and standard deviation equal to 1. We then use such noisy data to estimate the MIR of the different neural codes for different chemical coupling strengths and noise strengths $\sigma$. The dynamics is chaotic and comes from the deterministic system in Eqs. (\ref{HR_model_Nneurons}).

\begin{figure*}[!ht]
\centering{
\includegraphics[width=\textwidth,height=19cm]{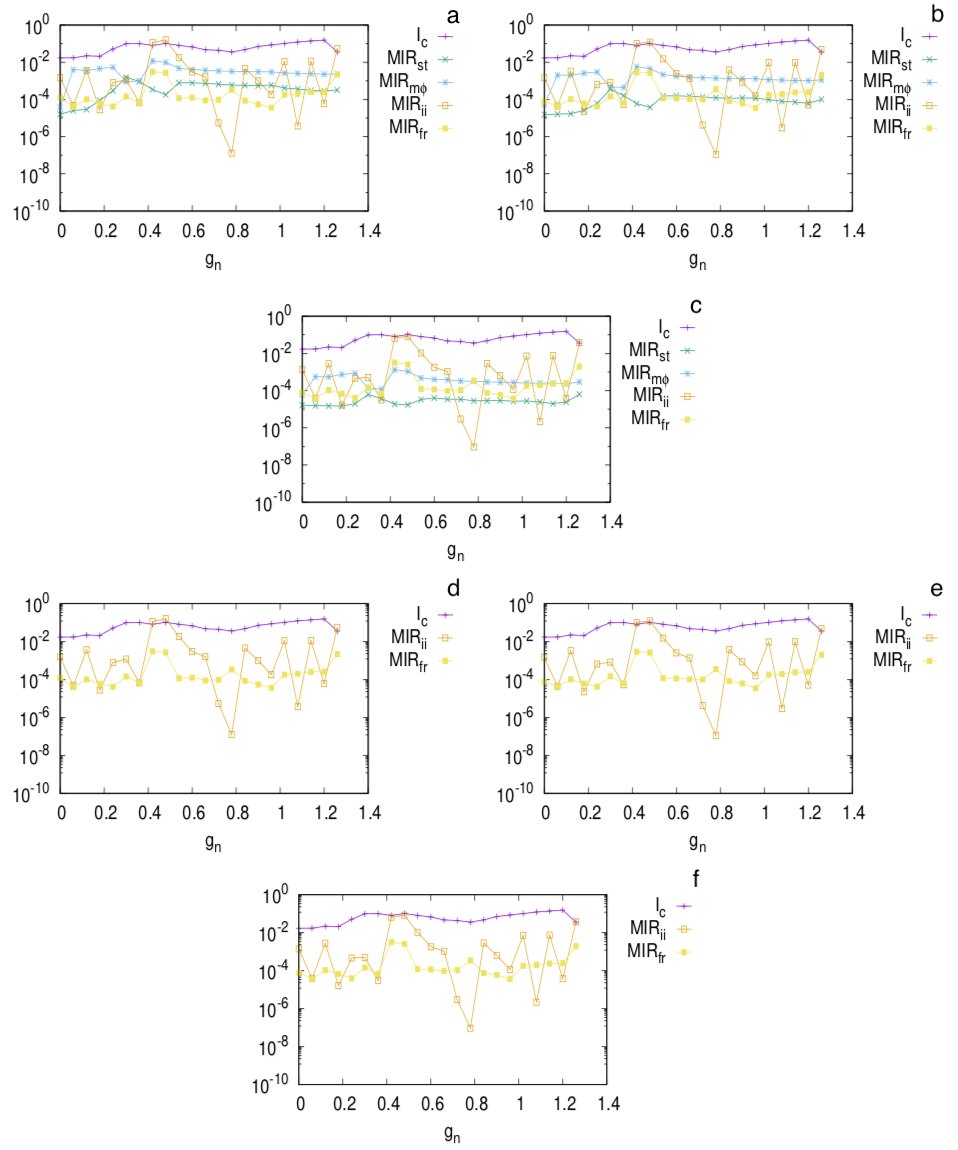}
}
\caption{\textbf{Results for the neural code used between two chemically connected noisy HR neurons.} Panel a: the MIR values of the different neural codes for noise strength $\sigma=0.4$. Panel b is for $\sigma=0.8$ and panel c for $\sigma=1.5$. Panels d, e and f are similar to a, b and c for MIR$_{ii}$ and MIR$_{fr}$ only and same $\sigma=0.4$ (panel d), $\sigma=0.8$ (panel e) and $\sigma=1.5$ (panel f) noise strengths as in the first three panels. We also plot $I_c$ in all panels to guide the eye. Notice that MIR$_{ii}$ and MIR$_{fr}$ in panels d, e and f, remain unaffected by the increase of the noise strength.}\label{fig_results_2noisyneurons}
\end{figure*}

We demonstrate these results in Fig. \ref{fig_results_2noisyneurons}. Particularly, we plot the MIR between the two neurons in Fig. \ref{fig_results_2neurons}a, for different chemical couplings and three noise strengths. Figure \ref{fig_results_2noisyneurons}a shows the same MIR quantities of Fig. \ref{fig_results_2neurons}b but for $\sigma=0.4$, panel b for $\sigma=0.8$ and panel c for $\sigma=1.5$. As $\sigma$ increases from zero, all MIR quantities start decreasing, except MIR$_{fr}$ and MIR$_{ii}$, which remain practically unaffected by the increase of noise strength. Figure \ref{fig_results_2noisyneurons} reveals that even though for small noise strengths, MIR$_{st}$ and MIR$_{m\phi}$ are larger than MIR$_{fr}$, they are nevertheless considerably affected by the increase of the noise strength. As we demonstrate in panels d, e and f, MIR$_{fr}$ and MIR$_{ii}$ prove to be consistently robust with respect to the increase of $\sigma$, even for values as high as 1.5. This underlines the importance of firing-rate against temporal codes, such as spike-timings or phase codes, which prove to be prone to noise contamination and to the transmission of smaller amounts of information per unit of time with the increase of noise strength.

Comparing Fig. \ref{fig_results_2neurons}b and Fig. \ref{fig_results_2noisyneurons}, it can be seen that in the presence of Gaussian additive noise (\ref{Gaussin_noise}), the various MIR quantities drop below $I_c$ around $g_n=0.48$. Also, the region where MIR$_{st} >$MIR$_ {m\phi}$ disappears. Particularly, with the noisy strength increasing, MIR$_{fr}$ and MIR$_{ii}$ became dominant as they are larger than MIR$_{m\phi}$ and MIR$_{st}$, respectively, except for some singular values. Our findings suggest that the firing-rate and interspike-intervals codes are more robust to readout noise.

\subsection{Neural Codes in a Communication System of Four Neurons}\label{subsec_4neurons}

\begin{figure*}[!ht]
\centering{
\includegraphics[width=\textwidth,height=14cm]{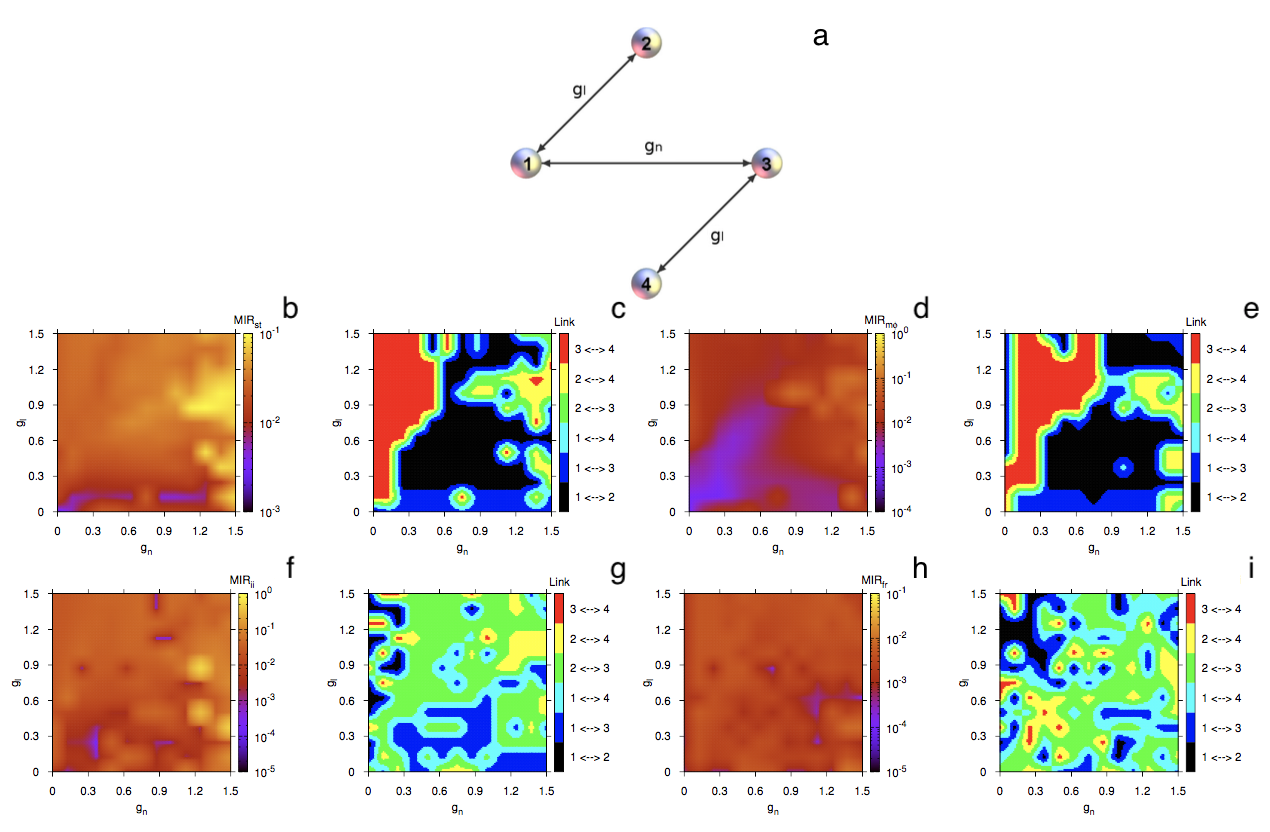}
}
\caption{\textbf{Topology and parameter spaces for the neural codes for four, non-noisy, HR neurons connected by 2 electrical and 1 chemical connection.} Panel a: the network of connections of the four neurons, where $g_n,g_l$ are the strengths of the chemical and electrical couplings, respectively. Panels b and c: the parameter spaces for MIR$_{st}$ for the two nodes that provide the largest MIR value and for the links that maximizes it, respectively. Panels d and e: similarly for MIR$_{m\phi}$. Panels f and g: similalry for MIR$_{ii}$. Panels h and i: similarly for MIR$_{fr}$. In all cases, the notation $i \leftrightarrow j$ indicates the bidirectional transfer of information between neurons $i$, $j$.}\label{fig_results_4neurons}
\end{figure*}

Here, we extend our study to the case of four bidirectionally connected non-noisy HR neurons, which are chemically and electrically coupled as shown in Fig. \ref{fig_results_4neurons}a. The first neuron is chemically connected with the third, whereas the first with the second and, the third with the fourth, are electrically connected. The strengths of the electrical and chemical connections are given by $g_l$ and $g_n$, respectively. The four neurons in Fig. \ref{fig_results_4neurons}a are arranged in a typical configuration when one wants to infer topology from information-theoretical quantities. The open-ring topology offers a way to test whether adjacent (non-adjacent) neurons share higher (lower) rates of information exchange. We aim to understand which neural code is best suited for the maximization of the rate of information-exchange for different coupling strengths and also, for which pairs this is so. The first and third neurons are the intermediates that facilitate the communication between the second and fourth. We consider the setup of Fig. \ref{fig_results_4neurons}a as a communication system in which, information is transmitted through the connections and reaches the neurons. In what follows and for each pair of coupling strengths, we estimate the MIR of the neural codes, and for each of them, we find its maximum MIR value and the corresponding pair of neurons that produces it. Then, for each coupling pair, we plot that maximum value and the corresponding pair of neurons.

In the following, we study the four neural codes for the model of four non-noisy neurons in Fig. \ref{fig_results_4neurons}a. In panels b and c, we plot the parameter spaces $(g_n,g_l)$ for the MIR$_{st}$ of the spike-timings and for the links that maximize it, respectively. The orange spots in panel b correspond to couplings that produce the largest amounts of MIR$_{st}$ whereas blue  to regions with the smallest MIR$_{st}$. The former occurs for relatively big chemical and electrical couplings whereas the latter for very small electrical and, small to large chemical couplings. Panel c reveals that, depending on the couplings, the largest amounts of MIR$_{st}$ are transmitted between different pairs of neurons, giving rise to a complicated pattern in the parameter space (see Fig. \ref{fig_results_4neurons}). The pattern is mainly characterized by the pair of neurons 3,4 (red) for small chemical and small to large electrical coupling strengths, by pair 1,2 (black) for comparatively small to large chemical and small to large electrical coupling strengths, and by many smaller-sized regions of different colors, such as blue, magenta, green and yellow that correspond to the remaining pairs of neurons.

We decided to use as a ``clock'' the first neuron as it is one of the two mediators that facilitate the transmission of information in this network (with the other one being neuron 3). This choice however is relative in the sense that for every pair of neurons we want to estimate MI, we should choose a ``clock''. In this sense, there is no universal ``clock'', but several ones can be used. This choice also intends to maximize the amount of MI measured between any two pairs of neurons.

The parameter space for MIR$_{m\phi}$ in panel d is mainly dominated by red  (that corresponds to comparatively large values), a smaller blue region of moderately very low values and a smaller orange region, for high chemical and electrical couplings, that corresponds to the highest observed MIR$_{m\phi}$ values in the parameter space. Similarly to panel c (for MIR$_{st}$), panel e for the pairs of neurons that maximize MIR$_{m\phi}$ shows that, depending on the coupling values, the largest amounts of MIR$_{m\phi}$ are transmitted between different pairs of neurons, giving rise to a complicated pattern in the parameter space, dominated mainly by the pair of neurons 3,4 (red) for small chemical and small to large electrical coupling strengths, by pair 1,2 (black) for comparatively small to large chemical and small to large electrical coupling strengths, and by many smaller-sized regions of different colors, (i.e. blue, magenta, green and yellow) that correspond to the remaining pairs of neurons.

The situation changes slightly in panel f for MIR$_{ii}$ where almost all the parameter space is dominated by red (of moderately large MIR$_{ii}$ values) with a few orange spots (very large values) and blue spots (of very low MIR$_{ii}$ values). The blue regions are considerably smaller in size than the blue region in panel d. The case for MIR$_{ii}$ is also different with respect to the pairs of neurons for which it is maximal. The parameter space in panel g reveals completely different structural properties than in panels c and e. Interestingly, the largest amounts of MIR$_{ii}$ occur for all pairs except 3,4 and, less for 1,2, implying that the first and third neurons play mainly the role of the facilitators in the transmission of information in the system.

A similar situation is happening for MIR$_{fr}$, with the parameter space in panel h looking uniformly covered by red  of moderately high MIR$_{fr}$ values and with a few quite small blue spots of very low values. MIR$_{fr}$ is less dependent on the coupling strengths. The parameter space for the links that maximize this quantity looks quite similar to that of MIR$_{ii}$, in the sense that the largest amounts of MIR$_{fr}$ occur for all pairs except 3,4 (red) and, less between 1 and 2 (black). This implies again that the first and third neurons play mainly the role of the facilitators in the transmission of information in the system.

A comparison of the parameter spaces in Fig. \ref{fig_results_4neurons} shows that the highest rate of information exchange can be attributed to the neural codes of the maximum points of the phase MIR$_{m\phi}$ and to the interspike intervals, MIR$_{ii}$. Moreover, MIR$_{fr}$ is practically unaffected by the coupling strengths, even though its maximum values are smaller than the maximum values of the neural codes based on the maximum points of the phase and interspike intervals. This result is in agreement with its performance in the case of the two neurons in Sec. \ref{subsec_2neurons}, where it attained the lowest values of all other codes. Interestingly, the pair of nodes more likely to exchange the largest amount of information per unit of time using the interspike-intervals and firing-rate codes are not adjacent in the network, whereas the spike-timings and the phase codes promote large exchange of information from adjacent nodes in the network. This provides evidence for the non-local character of firing-rate codes and local character of precise, spike-timings, codes.

\begin{figure*}[!ht]
\centering{
\includegraphics[width=\textwidth,height=14cm]{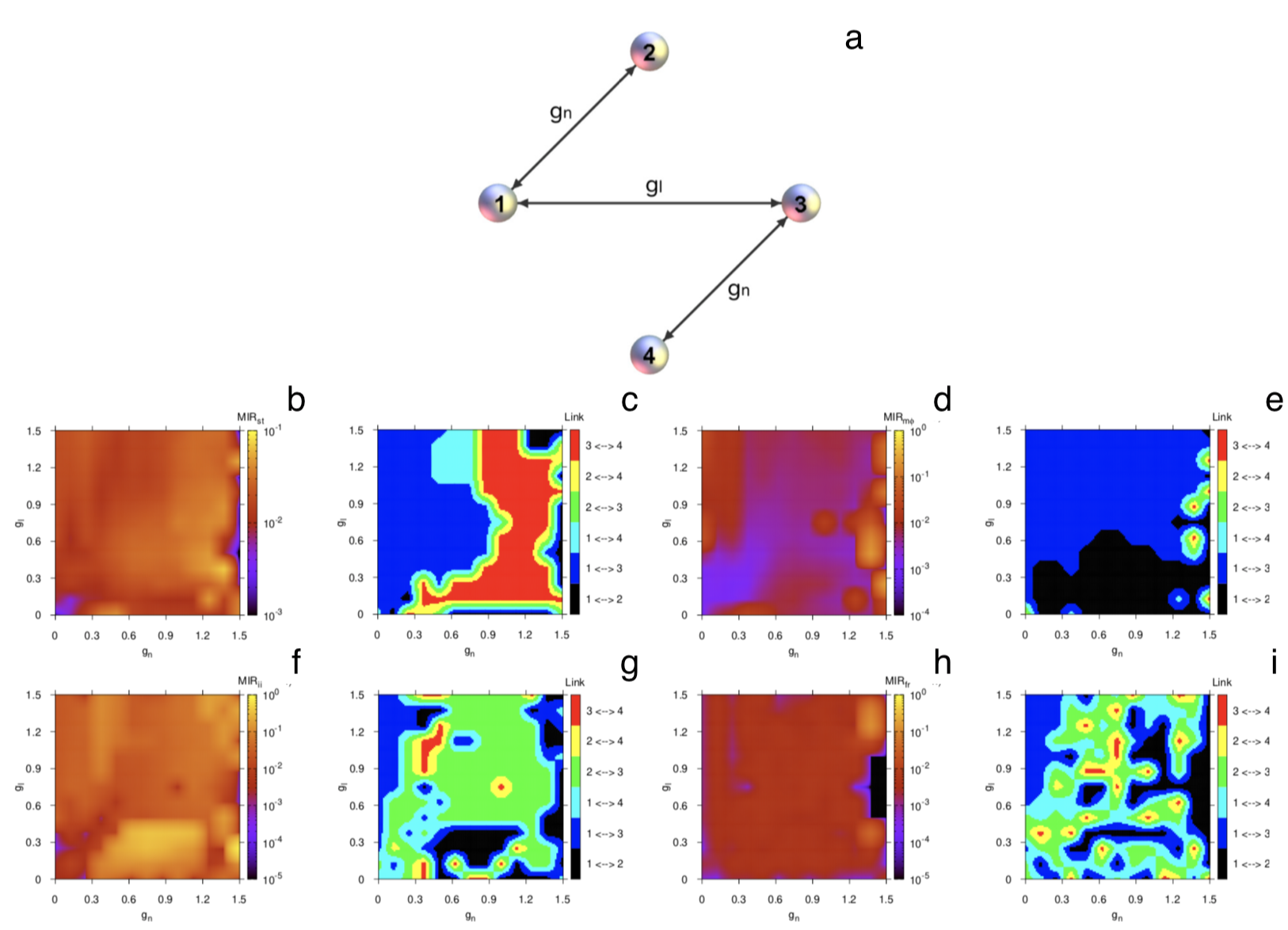}
}
\caption{\textbf{Topology and parameter spaces for the neural codes for four, non-noisy, HR neurons connected by 2 chemical and 1 electrical connection.} Panel a: the network of connections of the four neurons, where $g_n,g_l$ are the strengths of the chemical and electrical couplings, respectively. Panels b and c: the parameter spaces for MIR$_{st}$ for the two nodes that provide the largest MIR value and for the links that maximizes it, respectively. Panels d and e: similarly for MIR$_{m\phi}$. Panels f and g: similalry for MIR$_{ii}$. Panels h and i: similarly for MIR$_{fr}$. In all cases, the notation $i \leftrightarrow j$ indicates the bidirectional transfer of information between neurons $i$ and $j$.}\label{fig_results_4neurons_reviewer_question}
\end{figure*}

The latter result on the character of the codes is also backed by the results in Fig. \ref{fig_results_4neurons_reviewer_question} where the role of chemical and electrical connections has been swapped (compare Figs. \ref{fig_results_4neurons}a and \ref{fig_results_4neurons_reviewer_question}a). In particular, comparing panels c, e and g, i in Fig. \ref{fig_results_4neurons_reviewer_question}, one can again deduce that temporal codes (MIR$_{m\phi}$ and MIR$_{ii}$) perform optimally for adjacent neurons in the network whereas MIR$_{fr}$ and MIR$_{ii}$ for non-adjacent neurons. It becomes thus clear that the type of neural code with largest information transmission rate depends on network adjacency.


\subsection{Neural Codes in a Network of Twenty Neurons in a Bottleneck Configuration}\label{subsec_20neurons_bottleneck}

\begin{figure*}[!ht]
\centering{
\includegraphics[width=\textwidth,height=10.5cm]{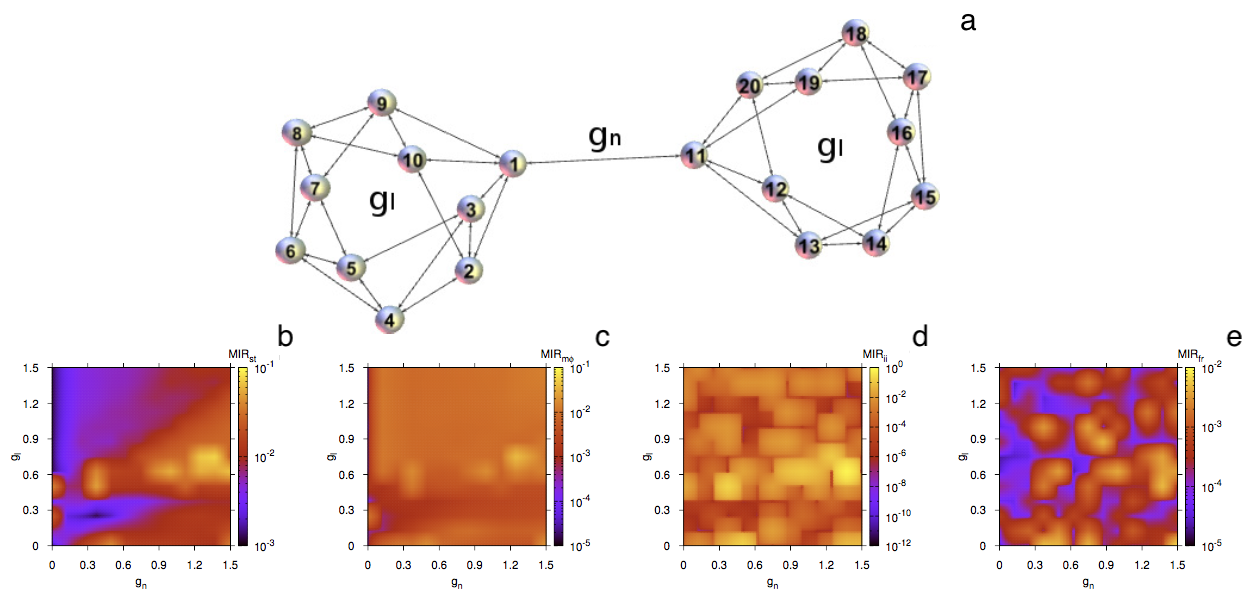}
}
\caption{\textbf{Topology with a bottleneck configuration and parameter spaces for the neural codes between two identical small-world, chemically connected, non-noisy clusters.} Panel a: the two identical clusters of electrically connected neurons with coupling strength $g_l$ and chemical strength $g_n$. Panel b: the parameter space for MIR$_{st}$, Panel c: similarly for MIR$_{m\phi}$, Panel d: similarly for MIR$_{ii}$ and panel e: for MIR$_{fr}$. The colors indicate the maximal MIR value that any two nodes exchange using a particular neural code.}\label{fig_results_20neurons_bottleneck}
\end{figure*}

Finally, we study the neural codes in an extended model of two identical clusters of 10 HR, non-noisy, neurons each. For simplicity, both clusters have the same small-world structure \cite{Wattsetal1998} and their neurons are internally coupled with electrical connections of strength $g_l$. This construction is interesting as it resembles a bottleneck, in which the two clusters communicate via the only link between the first and the eleventh neuron in the two clusters. The bottleneck is represented by a single, chemical link with strength $g_n$ that connects the two clusters. We used one interconnection as this is the simplest case in which information travels from one cluster to the other through the only chemical link. Moreover, it allows to draw interesting conclusions with regard to the neural codes for different coupling strengths. The topology in Fig. \ref{fig_results_20neurons_bottleneck}a is an example of how two neural networks would interact via a connection which implements a bottleneck. Again, the network is undirected and for each pair of coupling strengths, we estimate the MIR of the four neural codes. For each code, we find its maximum MIR and the corresponding pair of neurons that produces it. Then, for each coupling pair, we plot the maximum value. The network in Fig. \ref{fig_results_20neurons_bottleneck}a is motivated by the modular organization of the brain in which neurons are linked together to perform certain tasks and cognitive functions, such as pattern recognition, data processing, etc. Modular processors have to be sufficiently isolated and dynamically differentiated to achieve independent computations, but also globally connected to be integrated in coherent functions \cite{Zamora2010,Meunieretal2010,reimann2017cliques}. The structure in Fig. \ref{fig_results_20neurons_bottleneck}a helps us understand which neural code in modular neural networks is best suited for the transmission of the largest amount of information per unit of time and for which coupling strengths it occurs. Again, we treat the model in Fig. \ref{fig_results_20neurons_bottleneck}a as a communication system in which, information is transmitted through the links and reaches out to its different parts.

In Fig. \ref{fig_results_20neurons_bottleneck}a, we study the four neural codes. Panels b and c show the parameter space $(g_n,g_l)$ for MIR$_{st}$ and MIR$_{m\phi}$, respectively. Orange corresponds to couplings that produce the largest amounts of MIR values whereas blue or black to regions with the smallest values. Red is for intermediate MIR values. Panel b is for MIR$_{st}$ and reveals that the highest values can be achieved for large chemical and intermediate electrical coupling strengths. For example, for zero chemical coupling (i.e. $g_n=0$), MIR$_{st}$ is considerably smaller than for $g_n$ around 1.4. This underlines the importance of chemical connections among the clusters as they help the system transmit larger rates of information when neurons exchange information by the precise spike-timings (temporal code). In contrast, MIR$_{m\phi}$ seems to perform more consistently in the sense that the parameter space in panel c is more uniformly red with a few orange spots of large MIR values. Interestingly, this quantity becomes maximal for large chemical and moderate electrical coupling strengths, similarly to MIR$_{st}$. The situation is similar for MIR$_{ii}$, where again it becomes maximal for large chemical and moderate electrical coupling strengths. We note that the maximum MIR$_{ii}$ values of the orange spots in panel d are bigger by one order of magnitude than those in panels b and c. Finally, MIR$_{fr}$ still shows the same dependence on the coupling strengths to achieve its maximum, even though these maximum values are smaller by one or two orders of magnitude than those of the other three neural codes. Lastly, for MIR$_{fr}$, there are blue regions of very small values, distributed evenly in the parameter space.

Comparing the behavior of the various neural codes, the firing-rate seems to be less advantageous with respect to the maximum amounts of transmitted information for the rest. Our results suggest that it is more prominent for neurons to use temporal codes or the maximum points of their phases to communicate the maximal rate of information in modular neural networks, for chemical coupling strengths twice as that of the electrical coupling.

\section{Discussion}

In this paper we sought to study how information is encoded in neural activity as it is crucial for understanding the computations underlying brain functions. Information is encoded by patterns of activity within neural populations responsible for similar functions and the interest in studying them is related to how the ``neural code'' can be read, mainly to understand how the brain processes information to accomplish behavior and cognitive functions. Thus, investigating the fundamental properties of neural coding in networks of spiking neurons may allow for the interpretation of population activity and, for understanding better the limitations and abilities of neural computations.
 
To this end, we studied numerically neural coding in small-size networks of chemically and electrically coupled Hindmarsh-Rose spiking neurons. We have introduced four codes and have quantified the rate of information exchange for each code. The quantity used to measure the level of information exchanged is the Mutual Information Rate. The latter is by definition a symmetric quantity and cannot be used to infer the directionality of information flow. Therefore, our analysis cannot infer the direction of information exchange, only its intensity. In the simplest case of pairs of spiking neurons we have found that they exchange the largest amount of information per unit of time by opting for a temporal code in which the time of each spike conveys information which is transmitted to the other participating neuron. Our findings suggest that the firing-rate and interspike-intervals codes are more robust to additive Gaussian white noise.

We have also studied four, chemically and electrically, coupled neurons and found that the largest rates of information exchange are attributed to the neural codes of maximum points of their phases and interspike intervals. In this network and in the absence of noise, pairs of nodes that are likely to exchange the largest amount of information per unit of time using the interspike-intervals and firing-rate codes are not adjacent in the network, whereas the spike-timings and phase codes promote large rate of information exchange for adjacent neurons in the network. This finding is also backed by similar results obtained for the same network with the role of chemical and electrical connections swapped. Our results provide evidence for the non-local character of firing-rate codes and local character of precise spike-timings, temporal, codes in modular dynamical networks of spiking neurons. It becomes thus clear that the type of neural code with largest information transmission rate depends on network adjacency. This result, if possible to extend to larger neural networks, would suggest that small microcircuits of fully connected neurons, also known as cliques \cite{reimann2017cliques}, would preferably exchange information using temporal codes (spike-timings and phase codes), whereas on the macroscopic scale, where typically there will be pairs of neurons not directly connected due to the brain's sparsity, the most efficient codes would be the firing-rate and interspike-intervals codes.

For a relatively larger network of 20 neurons arranged in two equal-size small-world modules that form a bottleneck, our work shows that neurons choose a temporal code or the maximum points of their phases to transmit the maximal rate of information for chemical coupling strengths twice as that of the electrical coupling.

Our estimations of the Mutual Information Rate are based on the symbolic encoding of trajectories, and thus, depending on the encoding, similar results can be obtained with the standard binary code \cite{Strongetal1998,Baptistaetal2006}. Particularly, if the chosen time-window for the binary code is close to the average interspike-inter\-vals, MIR$_{ii}$ would produce similar values with the binary code, as 1's would encode spikes and 0's would typically encode relaxation in the neural activity.

Another possibility would be to use refinements in the estimation of the Mutual Information Rate, aiming at obtaining its true value for each. These refinements would correspond to the search for a generating Markov partition of higher order as in \cite{Rubidoetal2018}. Since this is out of the scope of the present paper, we leave it for a future publication. In fact, we sought to study whether looking at the instantaneous spike-timings would provide less information than the codes based on the interspike intervals. Our decision was driven by the question: in a time-series of events, what does carry more information? A code based on the times between events, or a code based on the precise times of the occurrence of the events?

Here, we have used chemical and electrical synapses with identical coupling strengths among all model neurons. As such, it is a limited study of relatively simple dynamical model neurons, small-size networks and equal synaptic connectivity. This choice was made for simplification. A similar study using unequal coupling strengths and larger networks would allow for more general results and would add more value from a neurophysiological perspective.

Lastly, we have shown the importance of firing-rate and interspike-intervals codes against the spike-timings code and those based on phases. The latter codes prove to be more prone to noise contamination and to the transmission of smaller amounts of information per unit of time with the increase of noise intensity.

\section{Acknowledgments}
This work was performed using the Maxwell high performance and ICSMB computer clusters of the University of Aberdeen. All authors acknowledge financial support provided by the EPSRC Ref: EP/I032606/1 grant. C. G. A. contributed to this work while working at the University of Aberdeen and then, while working at the University of Essex.



\begin{thebibliography}{10}
\expandafter\ifx\csname url\endcsname\relax
  \def\url#1{\texttt{#1}}\fi
\expandafter\ifx\csname urlprefix\endcsname\relax\def\urlprefix{URL }\fi
\expandafter\ifx\csname href\endcsname\relax
  \def\href#1#2{#2} \def\path#1{#1}\fi

\bibitem{Herculano-Houzel2012}
S.~Herculano-Houzel, The remarkable, yet not extraordinary, human brain as a
  scaled-up primate brain and its associated cost, Proc. Natl. Acad. Sci. USA
  109 (2012) 10661.

\bibitem{Perkeletal1969}
D.~H. Perkel, T.~H. Bullock, Neural coding, Neurosciences researchs symposium
  summaries. The MIT Press 3 (1969) 405--527.

\bibitem{Schneidmanetal2006}
E.~Schneidman, M.~J. Berry, R.~Segev, W.~Bialek, Weak pairwise correlations
  imply strongly correlated network states in a neural population, Nature 440
  (2006) 1007--12.

\bibitem{reimann2017cliques}
M.~W. Reimann, M.~Nolte, M.~Scolamiero, K.~Turner, R.~Perin, G.~Chindemi,
  P.~D{\l}otko, R.~Levi, K.~Hess, H.~Markram, Cliques of neurons bound into
  cavities provide a missing link between structure and function, Frontiers in
  Computational Neuroscience 11 (2017) 48.

\bibitem{Sporns2011}
O.~Sporns, Networks of the brain, Cambridge MA: The MIT Press, 2011.

\bibitem{Kandeletal1991}
E.~Kandel, J.~Schwartz, T.~Jessell, Principles of Neural Science (third ed.),
  Elsevier/North-Holland, Amsterdam, London, New York, 1991.

\bibitem{Gerstneretal1997}
W.~Gerstner, A.~K. Kreiter, H.~Markram, A.~V.~M. Herz, Neural codes: Firing
  rates and beyond, Proc. Natl. Acad. Sci. USA 94 (1997) 12740--12741.

\bibitem{Bohte2004}
S.~M. Both, The evidence for neural information processing with precise
  spike-times: A survey, Natural Computing 3 (2004) 195--206.

\bibitem{DiLorenzoetal2013}
P.~M. DiLorenzo, J.~D. Victor, Spike timing: {Mechanisms} and function, CRC
  Press, 2013.

\bibitem{Orametal2002}
M.~Oram, D.~Xiao, B.~Dritschel, K.~Payne, The temporal resolution of neural
  codes: does response latency have a unique role?, Trans. R. Soc. Lond. B 357
  (2002) 987--1001.

\bibitem{Thorpeetal1996}
S.~Thorpe, F.~Fize, C.~Marlot, Speed of processing in the human visual system,
  Nature 381 (1996) 520--522.

\bibitem{Johanssonetal2004}
R.~Johansson, I.~Birznieks, Fist spikes in ensembles of human tactile afferents
  code complex spatial fingertip events, Nature Neurrosci. 7 (2004) 170--177.

\bibitem{Baptistaetal2010}
M.~S. Baptista, F.~M. Kakmeni, C.~Grebogi, Combined effect of chemical and
  electrical synapses in {Hindmarsh-Rose} neural networks on synchronization
  and the rate of information, Phys. Rev. E 82 (2010) 036203.

\bibitem{Antonopoulosetal2015}
C.~G. Antonopoulos, S.~Srivastava, E.~d. S.~S. Pinto, M.~S. Baptista, Do brain
  networks evolve by maximizing their information flow capacity?, PLoS Comput.
  Biol. 11~(8) (2015) e1004372.

\bibitem{Bianco-Martinezetal2016}
E.~Bianco-Martinez, N.~Rubido, C.~G. Antonopoulos, M.~S. Baptista, Successful
  network inference from time-series data using mutual information rate, Chaos:
  An Interdisciplinary Journal of Nonlinear Science 26~(4) (2016) 043102.

\bibitem{Hopfield1995}
J.~J. Hopfield, Pattern recognition computation using action potential timing
  for stimulus representation, Nature 376 (1995) 33.

\bibitem{Rubidoetal2018}
O.~N. Rubido, C.~Grebogi, M.~S. Baptista, Entropy-based generating {Markov}
  partitions for complex systems, Chaos (in press).

\bibitem{Hindmarshetal1984}
J.~L. Hindmarsh, R.~M. Rose, A model of neuronal bursting using three coupled
  first order differential equations, Proc. R. Soc. London Ser. B 221 (1984)
  87--102.

\bibitem{Pereiraetal2007A}
T.~Pereira, B.~M. S., J.~Kurths, Phase and average period of chaotic
  oscillations, Phys. Let. A 362 (2007) 159--165.

\bibitem{Pereiraetal2007B}
T.~Pereira, B.~M. S., J.~Kurths, General framework for phase synchronization
  through localized sets, Phys. Rev. E 75 (2007) 026216.

\bibitem{Shannon1948}
C.~E. Shannon, A mathematical theory of communication, The Bell System
  Technical Journal 27 (1948) 379.

\bibitem{Borstetal1999}
A.~Borst, F.~E. Theunissen, Information theory and neural coding, Nat. Neurosc.
  2 (1999) 947--957.

\bibitem{Wibraletal2014}
M.~Wibral, R.~Vicente, J.~T. Lizier, Directed information measures in
  {Neuroscience}, Springer-Verlag Berlin Heidelberg, 2014.

\bibitem{Baptistaetal2012}
M.~S. Baptista, R.~M. Rubinger, E.~R. Viana, J.~C. Sartorelli, U.~Parlitz,
  C.~Grebogi, Mutual information rate and bounds for it, PLoS One 7 (2012)
  10:e46745.

\bibitem{Antonopoulosetal2014}
C.~G. Antonopoulos, E.~Bianco-Martinez, M.~S. Baptista, Production and transfer
  of energy and information in {Hamiltonian} systems, PLoS ONE 9~(2) (2014)
  e89585.

\bibitem{Benettin1980}
G.~Benettin, L.~Galgani, A.~Giorgilli, J.-M. Strelcyn, Lyapunov characteristic
  exponents for smooth dynamical systems and for {Hamiltonian} systems: {A}
  method for computing all of them. {Part} 1: {Theory} and {Lyapunov}
  characteristic exponents for smooth dynamical systems and for {Hamiltonian}
  systems: {A} method for computing all of them. {Part} 2: {Numerical}
  application, Meccanica 15 (1980) 9--20, 21--30.

\bibitem{Kraskovetal2004}
A.~Kraskov, H.~St\"ogbauer, P.~Grassberger, Estimating mutual information,
  Phys. Rev. E 69 (2004) 066138.

\bibitem{Spornsetal2004}
O.~Sporns, D.~R. Chialvo, M.~Kaiser, C.~C. Hilgetag, Organization, development
  and function of complex brain networks, Trends in Cognitive Sciences 8 (2004)
  418--425.

\bibitem{Paulusetal2001}
M.~Paulus, V.~Komarek, T.~Prochazka, Z.~Hrncir, K.~Sterbova, Synchronization
  and information flow in {EEGs} of epileptic patients, IEEE Engineering in
  Medicine and Biology Magazine 20~(5) (2001) 65--71.

\bibitem{Paninski2003}
L.~Paninski, Estimation of entropy and mutual information, Neural Computation
  15~(6) (2003) 1191--1253.

\bibitem{Steueretal2002}
R.~Steuer, J.~Kurths, C.~O. Daub, J.~Weise, J.~Selbig, The mutual information:
  {Detecting} and evaluating dependencies between variables, Bioinformatics
  18~(suppl 2) (2002) S231--S240.

\bibitem{Dimitrovetal2001}
A.~G. Dimitrov, A.~A. Lazar, J.~D. Victor, Information theory in neuroscience,
  Journal of computational neuroscience 30 (2011) 1--5.

\bibitem{CausalityChaos2018}
E.~Bianco-Martinez, M.~S. Baptista, Space-time nature of causality, Chaos 28
  (2018) 075509.

\bibitem{Strongetal1998}
S.~P. Strong, R.~K\"oberle, R.~R. de~Ruyter~van Steveninck, W.~Bialek, Entropy
  and information in neural spike trains, Phys. Rev. Lett. 80 (1998) 197--200.

\bibitem{Baptistaetal2016}
M.~S. Baptista, R.~M. Szmoski, R.~F. Pereira, S.~E. d.~S. Pinto, Chaotic,
  informational and synchronous behaviour of multiplex networks, Scientific
  Reports 6 (2016) 22617.

\bibitem{Sevilla-Escobozaetal2016}
R.~Sevilla-Escoboza, I.~Sendi\~na Nadal, I.~Leyva, R.~Guti\'errez, J.~Buld\'u,
  S.~Boccaletti, Inter-layer synchronization in multiplex networks of identical
  layers, Chaos 26 (2016) 065304.

\bibitem{Leyvaetal2017}
I.~Leyva, R.~Sevilla-Escoboza, I.~Sendi\~na Nadal, R.~Guti\'errez, J.~M.
  Buld\'u, S.~Boccaletti, Inter-layer synchronization in non-identical
  multi-layer networks, Scientific Reports 7 (2017) 45475.

\bibitem{Antonopoulosetal2017}
C.~G. Antonopoulos, M.~S. Baptista, Maintaining extensivity in evolutionary
  multiplex networks, PLoS ONE 12 (2017) 1--14.

\bibitem{Kacslemma}
M.~Kac, On the notion of recurrence in discrete stochastic processes, Bull.
  Amer. Math. Soc. 53 (1947) 1002?1010.

\bibitem{Shadlen1994569}
M.~N. Shadlen, W.~T. Newsome, Noise, neural codes and cortical organization,
  Current Opinion in Neurobiology 4~(4) (1994) 569--579.

\bibitem{RIS_0}
R.~B. Stein, E.~R. Gossen, K.~E. Jones, {Neuronal variability: {Noise} or part
  of the signal?}, Nat. Rev. Neurosci. 6~(5) (2005) 389--397.

\bibitem{baptista2008complex}
M.~Baptista, F.~M. Kakmeni, G.~Del~Magno, M.~Hussein, How complex a complex
  network of equal nodes can be?, arXiv preprint arXiv:0805.3487.

\bibitem{baptista2008finding}
M.~S. Baptista, J.~X. De~Carvalho, M.~S. Hussein, Finding quasi-optimal network
  topologies for information transmission in active networks, PloS one 3~(10)
  (2008) e3479.

\bibitem{Wattsetal1998}
D.~J. Watts, S.~H. Strogatz, Collective dynamics of small-world networks,
  Nature 393 (1998) 440--442.

\bibitem{Zamora2010}
G.~Zamora-L\'opez, C.~S. Zhou, J.~Kurths, Cortical hubs form a module for
  multisensory integration on top of the hierarchy of cortical networks, Front.
  Neuroinform. 4 (2010) 1.

\bibitem{Meunieretal2010}
D.~Meunier, R.~Lambiotte, E.~T. Bullmore, Modular and hierarchically modular
  organization of brain networks, Front. Neurosci. 4 (2010) 200.

\bibitem{Baptistaetal2006}
M.~S. Baptista, C.~Grebogi, R.~K\"oberle, Dynamically multilayered visual
  system of the multifractal fly, Phys. Rev. Lett. 97 (2006) 178102.

\end{thebibliography}
\bibliographystyle{elsarticle-num}
\providecommand{\noopsort}[1]{}\providecommand{\singleletter}[1]{#1}%

\end{document}